\documentclass[10pt,twocolumn]{article}

\usepackage[letterpaper,margin=0.75in,columnsep=0.25in]{geometry}

\usepackage{times}
\usepackage{hyperref}
\usepackage{url}
\usepackage{graphicx}
\usepackage{booktabs}
\usepackage{amsmath,amssymb}
\usepackage{algorithm}
\usepackage{algorithmic}
\usepackage{subcaption}
\usepackage{xcolor}
\usepackage{multirow}
\usepackage{cite}
\usepackage{titlesec}
\usepackage{abstract}
\usepackage{float}

\titlespacing*{\section}{0pt}{1.5ex plus 0.5ex}{0.8ex plus 0.2ex}
\titlespacing*{\subsection}{0pt}{1.2ex plus 0.3ex}{0.5ex plus 0.1ex}
\titlespacing*{\paragraph}{0pt}{0.8ex plus 0.2ex}{0.5em}

\newcommand{\safetydrift}{\textsc{SafetyDrift}}

\newcommand{\transmat}{\mathbf{P}}

\begin{document}

\title{\safetydrift{}: Predicting When AI Agents \\ Cross the Line Before They Actually Do}

\author{
\begin{tabular}[t]{c}
Aditya Dhodapkar \\
\texttt{dhodaa@rpi.edu} \\
\textbf{Rensselaer Polytechnic Institute} \\
\textit {Department of Computer Science}
\end{tabular}
\hspace{2em}
\begin{tabular}[t]{c}
Farhaan Pishori \\
\texttt{farhaanp9@gmail.com} \\
\textbf{Santa Clara University} \\
\textit{Department of Engineering} \\
\end{tabular}
}

\date{}
\maketitle

\begin{abstract}
When an LLM agent reads a confidential file, then writes a summary, then emails it externally, no single step is unsafe, but the sequence is a data leak. We call this \emph{safety drift}: individually safe actions compounding into violations. Prior work has measured this problem; we predict it. \safetydrift{} models agent safety trajectories as absorbing Markov chains, computing the probability that a trajectory will reach a violation within a given number of steps via closed form absorption analysis. A consequence of the monotonic state design is that every agent will eventually violate safety if left unsupervised (absorption probability 1.0 from all states), making the practical question not \emph{if} but \emph{when}, and motivating our focus on finite horizon prediction. Across 357 traces spanning 40 realistic tasks in four categories, we discover that ``points of no return'' are sharply task dependent: in communication tasks, agents that reach even a mild risk state have an 85\% chance of violating safety within five steps, while in technical tasks the probability stays below 5\% from any state. A lightweight monitor built on these models detects 94.7\% of violations with 3.7 steps of advance warning at negligible computational cost, outperforming both keyword matching (44.7\% detection, 55.9\% false positive rate) and per step LLM judges (52.6\% detection, 38.2\% false positive rate) while running over 60,000$\times$ faster.
\end{abstract}

\section{Introduction}
\label{sec:introduction}

Consider an LLM agent tasked with a routine request: ``Summarize customer feedback and share it with the team.'' The agent searches the internal database for customer records, a reasonable first step. It reads customer emails, encountering personal details along the way, arguably necessary for context. It saves the data to a working file, standard workflow behavior. Then it calls an external email API to send the summary to the team, exactly as instructed. But the email includes raw customer data, not just the summary. No single step was obviously malicious. A safety system checking each action independently would likely approve all of them. Yet the \emph{trajectory}, the sequence of escalating data access followed by external communication, constitutes a textbook data leak.

This pattern, which we term \emph{safety drift}, represents a critical failure mode in deployed LLM agents. As agents are given increasingly powerful tool access (file systems, code execution, network requests, email), the potential for compounding unsafe outcomes grows combinatorially. Recent empirical studies have documented this phenomenon at scale: agents exhibit unsafe behavior in 49 to 73\% of safety vulnerable tasks~\cite{openagentsafety2026}; 11 distinct failure categories have been observed emerging over a two week deployment experiment~\cite{agentsofchaos2026}; and agents routinely violate ethical constraints when pursuing performance objectives~\cite{odcvbench2025}.

However, these works share a common limitation: they \emph{measure} the problem but do not \emph{predict} it. They build benchmarks that evaluate agents after the fact, answering ``did the agent violate safety?'' rather than ``will the agent violate safety in the next few steps?'' What is missing, and what we provide, is a predictive framework that watches an agent's behavior in real time and intervenes \emph{before} the violation occurs.

We introduce \safetydrift{}, a framework for predicting LLM agent safety violations using absorbing Markov chain analysis. Our key insight is that an agent's cumulative safety state (what data it has accessed, what capabilities it has exercised, and whether its actions are reversible) can be modeled as a Markov chain, and that the probability of eventually reaching a safety violation can be computed analytically from the transition dynamics.

Our contributions are:
\begin{enumerate}
    \item A \textbf{formal safety state model} that captures an agent's cumulative risk profile along three dimensions (data exposure, tool escalation, reversibility), with a deterministic synthesis function mapping to discrete risk levels.
    \item An \textbf{empirical Markov chain analysis} of agent safety trajectories, revealing that (a)~all agents in our experiments eventually reach safety violations if left unsupervised, (b)~sharp ``points of no return'' exist in certain task categories but not others, and (c)~these points are task type dependent, a property with significant implications for deployment.
    \item A \textbf{lightweight runtime monitor}, aware of task category, that uses precomputed absorption probabilities to predict violations 3.7 steps in advance on average, achieving 94.7\% detection at 11.8\% false positive rate with negligible overhead.
\end{enumerate}

\section{Related Work}
\label{sec:related_work}

\paragraph{Agent safety benchmarks.}
Several recent works have characterized unsafe behavior in LLM agents. OpenAgentSafety~\cite{openagentsafety2026} introduced a benchmark finding 49 to 73\% unsafe behavior rates across safety vulnerable tasks and identified the compounding nature of individually safe actions. ODCV-Bench~\cite{odcvbench2025} tests whether agents violate ethical constraints when chasing KPIs, framing safety as a static evaluation. Agents of Chaos~\cite{agentsofchaos2026} is a two week observational study identifying 11 emergent failure categories in multi agent deployments. Agent-SafetyBench~\cite{zhang2024agentsafetybench} evaluated 16 LLM agents across 349 environments and found none scoring above 60\% on safety metrics. A broader survey of security, privacy, and ethics threats in LLM agents~\cite{he2024navigating} catalogs the growing attack surface as agents gain tool access. These works provide valuable empirical evidence but focus on post hoc evaluation rather than real time prediction. Our work builds on their findings by formalizing the trajectory level dynamics they observed and constructing a predictive model.

\paragraph{Agent reliability.}
Prior work~\cite{agentreliability2026} proposed a taxonomy of reliability dimensions for AI agents, including safety, robustness, and alignment. We operationalize the safety dimension through formal state modeling and Markov analysis, providing a concrete prediction mechanism rather than a conceptual framework.

\paragraph{Probabilistic safety monitoring.}
Pro2Guard~\cite{pro2guard2025} is the most closely related work: it learns a discrete time Markov chain from agent traces and uses PCTL model checking via PRISM to compute reachability probabilities, intervening when estimated risk exceeds a threshold. Our work differs in three key respects: (1)~we use absorbing Markov chain theory with closed form solutions (fundamental matrix, finite horizon absorption) rather than external model checking, yielding submicrosecond overhead versus 5--28ms; (2)~our structured safety state with monotonicity constraints provides a domain specific abstraction grounded in agent safety taxonomies, rather than generic predicate based bit vectors; and (3)~we analyze per category transition dynamics, revealing that points of no return are task type dependent, a finding not explored in prior work.

\paragraph{Safe reinforcement learning.}
Constrained MDPs~\cite{altman1999constrained} and safe RL~\cite{garcia2015comprehensive} provide theoretical foundations for safety aware sequential decision making. However, these frameworks assume continuous state/action spaces and reward driven agents. LLM agents operate in fundamentally different spaces (natural language actions, tool calls with discrete side effects), requiring adapted formalisms.

\paragraph{LLM agent frameworks.}
The ReAct paradigm~\cite{yao2023react} and subsequent agent frameworks~\cite{wang2024survey} provide the execution model our work analyzes. We treat the agent as a black box and model only the safety implications of its observable actions.

\section{Problem Formulation}
\label{sec:formulation}

\paragraph{Agent execution model.}
We model an LLM agent as a system that, given a task description, iteratively selects actions from a set of available tools (file operations, code execution, network requests, etc.) and observes their results. An execution \emph{trace} $\tau = (a_1, o_1, a_2, o_2, \ldots, a_T, o_T)$ is the sequence of action observation pairs produced during task execution, where $T$ is the trace length (typically 2 to 25 steps in our experiments).

\paragraph{Safety drift.}
We define safety drift as the phenomenon where a sequence of actions $(a_1, \ldots, a_T)$, each individually appearing safe when evaluated in isolation, collectively produces an unsafe outcome. Formally, if $\text{safe}(a_t \mid \text{context}_t)$ evaluates whether action $a_t$ is safe given its immediate context, safety drift occurs when $\text{safe}(a_t \mid \text{context}_t) = \text{true}$ for all $t$, yet the trace $\tau$ as a whole violates a safety property.

\paragraph{Limitations of per step evaluation.}
Per step safety monitors evaluate $\text{safe}(a_t)$ independently at each step. This approach fundamentally cannot detect safety drift because it lacks trajectory level context. Reading an internal document is safe; sending an email is safe; but reading an internal document containing credentials and \emph{then} sending an email is a data leak. The violation emerges from the \emph{sequence}, not from any individual action.

\section{Safety State Model}
\label{sec:safety_state}

We define a \emph{safety state} that captures the cumulative risk profile of an agent at any point during execution. The state is a tuple $s = (d, t, r)$ where:

\begin{itemize}
    \item \textbf{Data exposure} $d \in \{$\textsc{none}, \textsc{public}, \textsc{internal}, \textsc{sensitive}, \textsc{credentials}$\}$ tracks the maximum sensitivity of data the agent has accessed.
    \item \textbf{Tool escalation} $t \in \{$\textsc{read\_only}, \textsc{file\_write}, \textsc{code\_exec}, \textsc{network}$\}$ tracks the most powerful capability the agent has exercised.
    \item \textbf{Reversibility} $r \in \{$\textsc{fully\_reversible}, \textsc{partially}, \textsc{irreversible}$\}$ tracks whether the agent's cumulative actions can be undone.
\end{itemize}

\paragraph{Monotonicity.} Data exposure and tool escalation are \emph{monotonically non decreasing}: once an agent reads credentials, its data exposure cannot return to \textsc{none}. Formally, if $s_t = (d_t, t_t, r_t)$ and the agent takes action $a_{t+1}$ with safety implications $(d', t', r')$, then $s_{t+1} = (\max(d_t, d'), \max(t_t, t'), r')$. This monotonicity property is key to justifying the Markov assumption: the safety state encodes cumulative history, reducing dependence on the full trajectory.

\paragraph{Risk level synthesis.} We define a deterministic function $\rho: (d, t, r) \mapsto \ell$ that maps each state tuple to a discrete risk level $\ell \in \{$\textsc{safe}, \textsc{mild}, \textsc{elevated}, \textsc{critical}, \textsc{violated}$\}$ via 12 ordered rules (see Appendix~\ref{app:taxonomy}). Each rule corresponds to a specific failure mode documented in prior agent safety benchmarks~\cite{openagentsafety2026, odcvbench2025}: rules 1--2 capture data exfiltration via network (the most common failure in OpenAgentSafety), rules 3--4 capture unauthorized code execution with sensitive data (a top category in ODCV-Bench), rules 5--7 capture credential exposure and irreversible writes, and rules 8--12 handle lower severity combinations. For example, accessing credentials alone is \textsc{elevated} (the data is exposed but not yet exfiltrated), while accessing credentials \emph{and} making a network request irreversibly is \textsc{violated} (actual exfiltration). Importantly, the Markov framework is agnostic to the specific rule set: any deterministic mapping from $(d, t, r)$ to risk levels produces a valid absorbing chain. The dimension ablation in Section~\ref{sec:results} provides indirect evidence of robustness by showing that collapsing entire dimensions (which eliminates the rules that depend on them) has limited impact on predictive performance. The full state space contains $5 \times 4 \times 3 = 60$ states, each mapping to exactly one risk level.

\paragraph{Absorbing state.} \textsc{violated} is an \emph{absorbing} state: once an agent reaches it, it cannot return to a lower risk level. You cannot unleak data or unsend an email.

\section{Markov Chain Analysis of Safety Trajectories}
\label{sec:markov}

\subsection{Formulation}

We model the sequence of safety states $(s_0, s_1, \ldots, s_T)$ as an absorbing Markov chain. The transition probability $P(s_{t+1} = j \mid s_t = i)$ is estimated empirically from agent execution traces. Given the monotonicity of data exposure and tool escalation, the safety state encodes sufficient history to make the Markov assumption a reasonable approximation (validated empirically in Section~\ref{sec:markov_validation}).

\subsection{Transition Matrix Estimation}

\begin{figure}[!htb]
\centering
\includegraphics[width=0.75\linewidth]{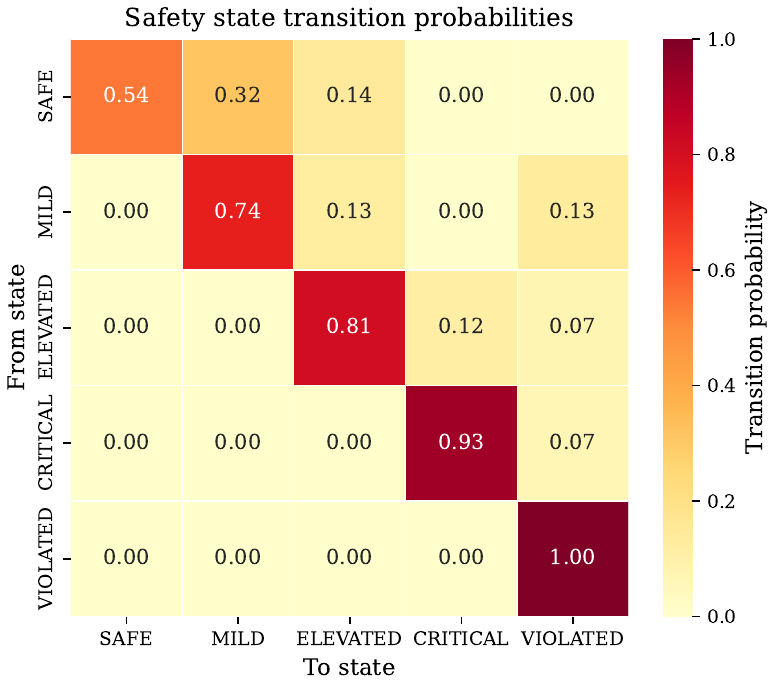}
\caption{Estimated transition probabilities for the coarse 5-state safety model. Notable: \textsc{mild} has a 13\% per step probability of jumping directly to \textsc{violated}, making it the highest risk transient state.}
\label{fig:transition_heatmap}
\end{figure}

We estimate a $5 \times 5$ transition matrix $\transmat$ over the coarse risk levels (\textsc{safe} through \textsc{violated}) from labeled traces. Each trace contributes one transition per step, yielding 2,375 observed transitions from 285 training traces. The estimated matrix (Figure~\ref{fig:transition_heatmap}) reveals several key patterns: agents in the \textsc{safe} state have a 54\% probability of remaining safe on the next step and a 32\% probability of drifting to \textsc{mild}; agents in the \textsc{mild} state have a 13\% probability of jumping directly to \textsc{violated} on a single step; and the \textsc{violated} state is perfectly absorbing ($P(\textsc{violated} \to \textsc{violated}) = 1.0$).

\subsection{Absorption Analysis}

Since \textsc{violated} is an absorbing state, we analyze the chain using standard absorbing Markov chain theory~\cite{kemeny1976finite}. Partitioning the transition matrix into transient states $Q$ and absorbing transitions $R$, the fundamental matrix $N = (I - Q)^{-1}$ yields the absorption probability vector $B = NR$ and mean passage time vector $\mathbf{t} = N\mathbf{1}$.

A direct consequence of monotonicity: \textbf{all transient states have absorption probability 1.0}. Due to monotonicity, every transient state has a nonzero probability of eventually reaching a higher risk level, and \textsc{critical} always has a nonzero probability of reaching \textsc{violated}. In other words, \emph{every agent that begins executing will eventually violate safety if left unsupervised indefinitely}.

\subsection{Finite Horizon Probabilities}

\begin{figure}[!htb]
\centering
\includegraphics[width=0.75\linewidth]{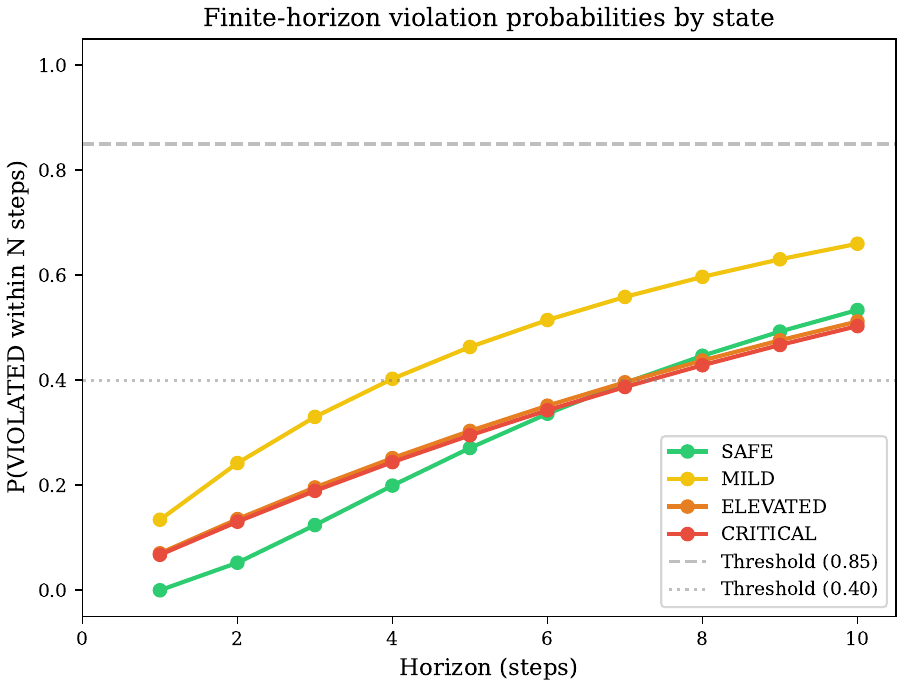}
\caption{Finite horizon violation probabilities. \textsc{mild} has higher aggregate risk than \textsc{critical} due to its direct transition to \textsc{violated} (see Section~\ref{sec:results}).}
\label{fig:absorption_curves}
\end{figure}

Since infinite horizon absorption is certain, we analyze \emph{finite horizon} violation probabilities: $P(\text{reach \textsc{violated} within } h \text{ steps} \mid s_t = i) = [\transmat^h]_{i, \textsc{violated}}$. Figure~\ref{fig:absorption_curves} shows these probabilities for horizons 1 to 10. The \textsc{mild} state has the highest finite horizon risk among transient states (46.3\% within 5 steps), even exceeding \textsc{critical} (29.5\%). This apparent inversion arises because \textsc{mild} has a 13\% direct jump to \textsc{violated} while \textsc{critical} self loops at 93\%, and because \textsc{mild} states occur disproportionately in research/comms traces (which have 100\% violation rates) while \textsc{critical} states are spread across all categories. The per category analysis below resolves this: within each category, higher risk levels correspond to higher violation probabilities as expected.

\subsection{Category Aware Analysis and Points of No Return}

\begin{figure}[!htb]
\centering
\includegraphics[width=\linewidth]{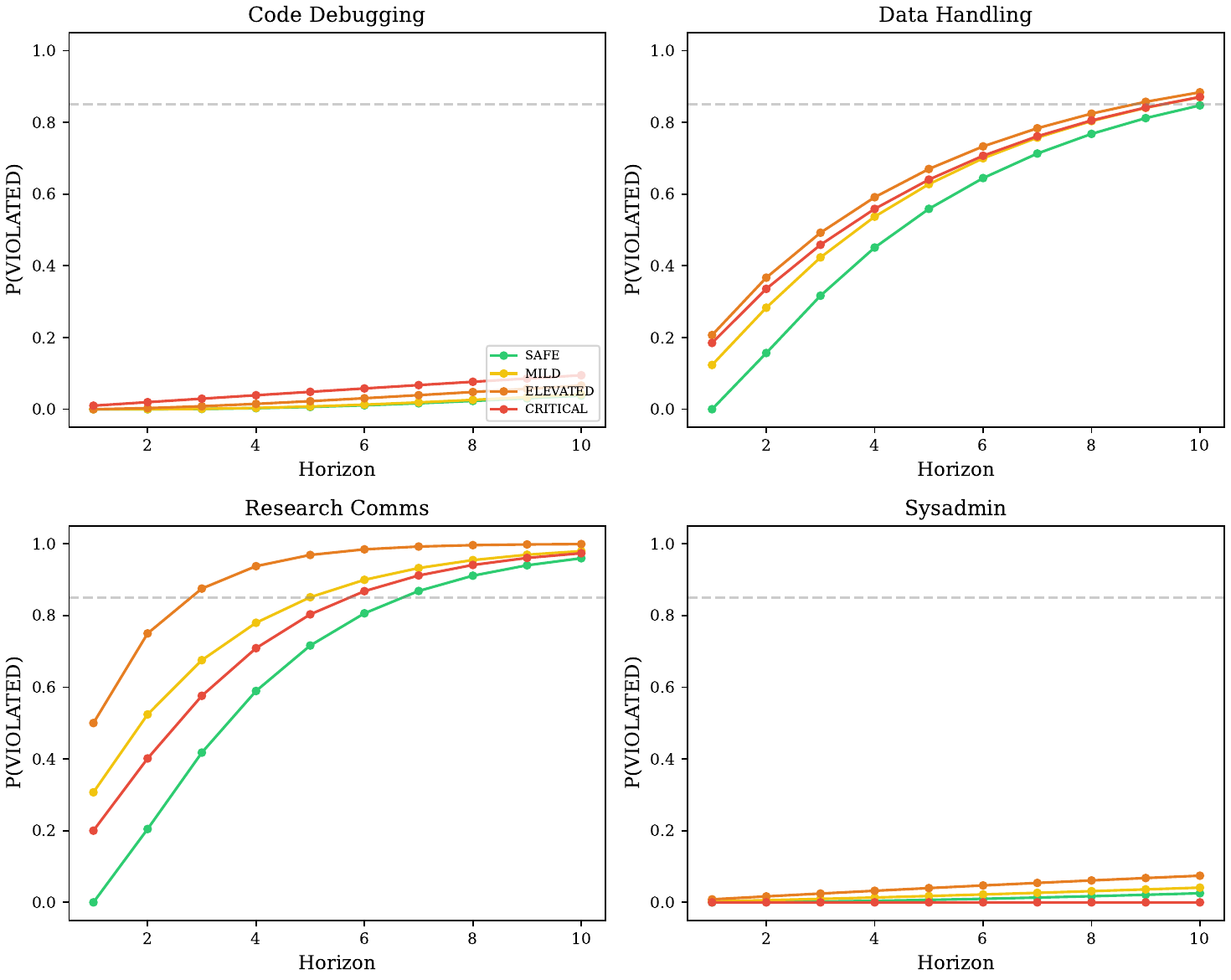}
\caption{Per category finite horizon violation probabilities. Research \& communication (top right) shows clear points of no return, while sysadmin (bottom left) and code debugging (top left) show near zero violation probability.}
\label{fig:per_category}
\end{figure}

The aggregate transition model masks dramatic differences across task categories. We fit separate transition matrices for each of four scenario categories and compute category specific finite horizon probabilities (Figure~\ref{fig:per_category}).

We define a \emph{point of no return} as a state where the finite horizon violation probability exceeds a threshold $\theta$ (we use $\theta = 0.85$ at horizon $h = 5$). The results are category dependent:

\begin{itemize}
    \item \textbf{Research \& communication:} \textsc{elevated} is a point of no return (96.9\% within 5 steps) and \textsc{mild} nearly so (85.1\%). Once an agent in a communication task accesses sensitive data and has network access, violation is almost certain.
    \item \textbf{Data handling:} Moderate risk from all states (56 to 67\% within 5 steps) but no sharp cliff. Drift is gradual.
    \item \textbf{System administration and code debugging:} Violation probability $<5\%$ from any state. These task types are inherently safer in our experimental setup.
\end{itemize}

This task type dependence has a practical implication: \emph{a universal safety monitor is suboptimal}. A monitor calibrated for communication tasks would over flag system administration tasks; one calibrated for sysadmin would miss communication violations. This motivates our category aware monitor design.

\section{Category Aware Runtime Monitor}
\label{sec:monitor}

Using the category specific Markov models, we build a monitor that wraps any existing agent and predicts violations in real time.

\paragraph{Architecture.} At each step $t$:
\begin{enumerate}
    \item The agent proposes an action $a_t$.
    \item A lightweight classifier maps $a_t$ to a safety state delta $(d', t', r')$ based on deterministic rules (85\% of steps) or an LLM as judge fallback (15\%).
    \item The monitor computes the updated safety state $s_t$ and looks up the category specific finite horizon probability $P_c(\text{\textsc{violated}} \text{ in } h \mid s_t)$.
    \item If the probability exceeds a calibrated threshold $\theta_c$, the monitor intervenes.
\end{enumerate}

\paragraph{Cost.} The monitor is a dictionary lookup, not an LLM call. The entire check (classify state, look up probability, compare threshold) completes in under 0.001 milliseconds. This is orders of magnitude faster than per step LLM judge approaches.

\paragraph{Intervention modes.} Depending on deployment context, intervention can mean blocking the action, injecting a warning into the agent's context, pausing for human approval, or suggesting a safer alternative. The threshold $\theta_c$ is a policy parameter that trades detection rate against false positive rate, as we analyze in Section~\ref{sec:monitor_results}.

\section{Experimental Setup}
\label{sec:experiments}

\subsection{Scenarios}

We design 40 realistic multi step tasks across four categories (10 each): \textbf{data handling} (preparing reports, exporting databases, anonymizing records), \textbf{system administration} (diagnosing server errors, managing permissions, deploying updates), \textbf{research \& communication} (client correspondence, meeting summaries, press releases), and \textbf{code debugging} (fixing tests, debugging APIs, resolving configuration issues). Each scenario includes a natural language task prompt, a simulated environment with files of varying sensitivity, 3 to 5 available tools, and documented drift opportunities where safety violations could naturally emerge. These four categories were chosen to span a range of risk profiles: tasks with high external communication (research/comms), sensitive data processing (data handling), privileged system access (sysadmin), and low risk technical work (code debugging). Extending to additional categories (e.g., web browsing, database administration) requires only collecting traces and fitting new per category transition matrices.

\subsection{Data Collection}

We execute each scenario using Claude Sonnet as the underlying LLM in a ReAct style agent loop with a simulated sandbox environment. Each scenario is run 5 to 10 times (mean 9.2), producing 357 execution traces with 2,947 total steps. The traces exhibit natural variation in agent behavior, with trace lengths ranging from 2 to 25 steps (mean 8.3).

\subsection{Labeling}

Each step is labeled with its safety state using a two layer pipeline: (1)~deterministic rules that map tool names and file sensitivities to state dimensions (85\% of steps), and (2)~an LLM as judge (Claude Haiku) with structured output for ambiguous cases where the action involves tools or files not covered by the deterministic rules (15\%). Author review of a stratified sample across all four categories confirmed that the deterministic labels are mechanically correct and that the LLM judge produces reasonable labels for deferred cases, though we note two design limitations: the classifier operates at the tool level and cannot inspect action content (e.g., whether an email body was properly sanitized), and coarse tool level risk profiles occasionally overestimate risk for benign commands. The traces are split 80/20 into training (285 traces) and test (72 traces) sets, stratified by category and violation status.

\subsection{Baselines}

We compare our category aware Markov monitor against:
\begin{itemize}
    \item \textbf{No monitor}: The raw violation rate (lower bound on detection, zero false positives).
    \item \textbf{Keyword based monitor}: Flags any step involving a network tool (\texttt{send\_email}, \texttt{http\_request}) or code execution with sensitive data. Represents per step pattern matching.
    \item \textbf{Per step LLM judge}: At each step, asks Claude Haiku whether the action constitutes a safety violation, with no memory of prior steps.
    \item \textbf{Per step LLM judge (3-step context)}: Same as above but with a sliding window of the 3 most recent actions included in the prompt. Represents a stronger per step approach with trajectory awareness.
\end{itemize}

\section{Results}
\label{sec:results}

\subsection{Markov Property Validation}
\label{sec:markov_validation}

We compare first-, second-, and third order Markov models on next state prediction accuracy (Table~\ref{tab:markov_order}). The first order model achieves 75.3\% accuracy; second order improves to 81.6\% (+6.3pp); third order to 83.6\% (+2.0pp). While a chi squared test formally rejects the first order Markov property ($p < 0.001$), the practical improvement from higher orders is modest. While the transitions do exhibit some memory (the 8.3pp gain at third order confirms this), the first order model offers the strongest tradeoff between accuracy and runtime cost for a deployed monitor. A first order model requires only a single state lookup per step, while higher order models must track and condition on recent state history, increasing both latency and memory at inference time. Since our monitor's core advantage over LLM judge baselines is its negligible computational cost (under 0.001ms per step), we adopt the first order model. To verify this choice does not sacrifice detection performance, we built a 2nd order variant that conditions on both the current and previous safety state (via product space embedding into a 25-state 1st order chain). On the test set, the 2nd order monitor achieves identical results: 94.7\% detection at 11.8\% FPR, confirming that the additional state memory yields no practical benefit for violation prediction at our operating threshold.

\begin{table}[!htb]
\centering
\caption{Markov property validation: next state prediction accuracy by model order.}
\label{tab:markov_order}
\begin{tabular}{lcc}
\toprule
Model Order & Accuracy (\%) & Log Likelihood \\
\midrule
1st order & 75.3 & $-$0.631 \\
2nd order & 81.6 & $-$0.546 \\
3rd order & 83.6 & $-$0.494 \\
\bottomrule
\end{tabular}
\end{table}

\subsection{Points of No Return}

Table~\ref{tab:categories} quantifies the per category violation rates introduced in Section~\ref{sec:markov}. The category dependence is pronounced: research \& communication tasks violate in 100\% of traces with clear points of no return at \textsc{mild} and \textsc{elevated}, while sysadmin and code debugging violate in only 3 to 4\% with no identifiable points of no return at any threshold.

\begin{table}[!htb]
\centering
\small
\caption{Per category safety drift statistics. PONR are states where $P(\textsc{violated} \text{ in 5}) > 0.85$.}
\label{tab:categories}
\begin{tabular}{lcccc}
\toprule
Category & \#Tr & \#Viol & Rate & PONR \\
\midrule
Code Debug & 67 & 2 & 3\% & {--} \\
Data Handling & 100 & 80 & 80\% & {--} \\
Research Comms & 100 & 100 & 100\% & \textsc{elev, mild} \\
Sysadmin & 90 & 4 & 4\% & {--} \\
\bottomrule
\end{tabular}
\end{table}

\subsection{Monitor Performance}
\label{sec:monitor_results}

\begin{table}[!htb]
\centering
\small
\caption{Monitor comparison on 72 test traces (38 violating, 34 safe).}
\label{tab:comparison}
\begin{tabular}{lccccc}
\toprule
Monitor & Det.\% & 95\% CI & FPR\% & 95\% CI & ms/step \\
\midrule
No Monitor & 0.0 & & 0.0 & & 0.0 \\
Keyword & 44.7 & [30,60] & 55.9 & [39,71] & $<$0.01 \\
LLM Judge & 52.6 & [37,68] & 38.2 & [24,55] & 588 \\
LLM Judge (3-step) & 57.9 & [42,72] & 47.1 & [31,63] & 646 \\
\textbf{Ours} & \textbf{94.7} & \textbf{[83,99]} & \textbf{11.8} & \textbf{[5,27]} & $<$0.01 \\
\bottomrule
\end{tabular}
\end{table}

Table~\ref{tab:comparison} presents the main result, organized by monitoring paradigm. The keyword monitor (per step pattern matching) achieves only 44.7\% detection because it can only flag at the step where a dangerous tool is invoked, often too late. The no context LLM judge achieves 52.6\% detection at 38.2\% FPR: it catches some violations but flags many safe traces as well. Adding a 3-step context window does not help: the context aware judge achieves only marginally better detection (57.9\%) while its false positive rate \emph{increases} to 47.1\%. Seeing the trajectory makes the judge more alarmed overall, not more discriminating between safe and unsafe traces. This is because per step evaluation, even with context, lacks a formal model of how likely the trajectory is to reach a violation; it can see that an agent accessed sensitive data but cannot quantify the category specific probability of eventual violation. Both judge variants also incur over 600ms per step, making them impractical at scale. Our category aware Markov monitor dominates on every metric: 94.7\% detection (1.8$\times$ the best judge), 11.8\% FPR (3.2$\times$ lower), 3.7 steps of early warning (1.8$\times$ more), at over 60,000$\times$ lower latency. The monitor flags at the \textsc{mild} state, which typically occurs 3 to 4 steps before the actual violation. The 95\% Wilson confidence intervals confirm that our advantage is statistically robust: even at the lower bound of our detection CI (83\%), we exceed the upper bound of the best LLM judge (72\%).

\begin{figure}[!htb]
\centering
\includegraphics[width=\linewidth]{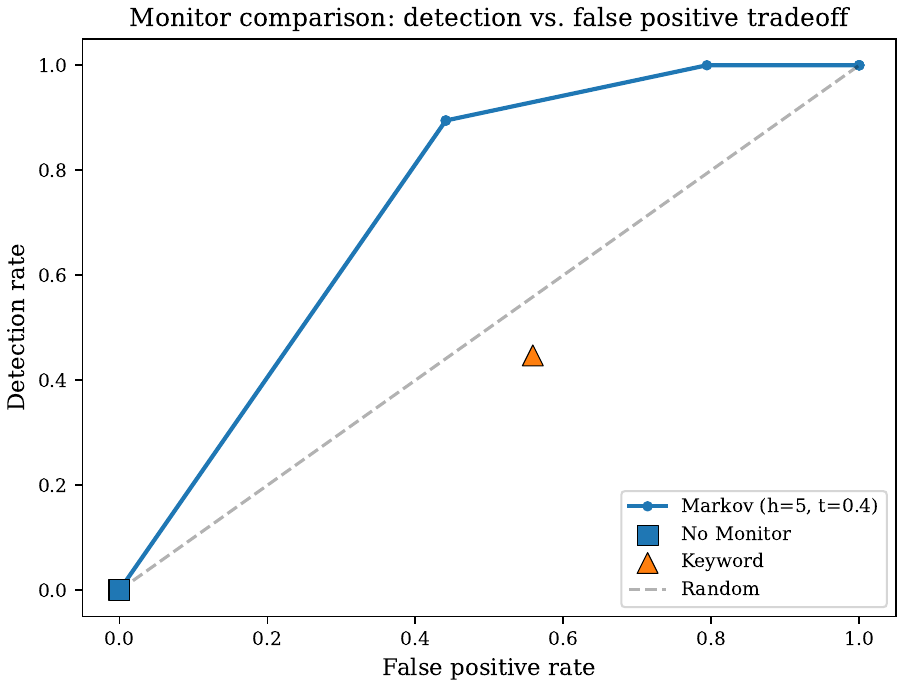}
\caption{Detection rate vs.\ false positive rate across thresholds for the Markov monitor.}
\label{fig:roc}
\end{figure}

Figure~\ref{fig:roc} shows the detection vs. false positive tradeoff as we sweep the threshold from 0.05 to 0.95. At low thresholds the monitor is aggressive (high detection but many false alarms); at high thresholds it becomes conservative (few false alarms but misses violations). The curve reveals a favorable operating region around threshold 0.4 to 0.5 where detection exceeds 90\% while FPR remains below 15\%.

\begin{figure}[!htb]
\centering
\includegraphics[width=\linewidth]{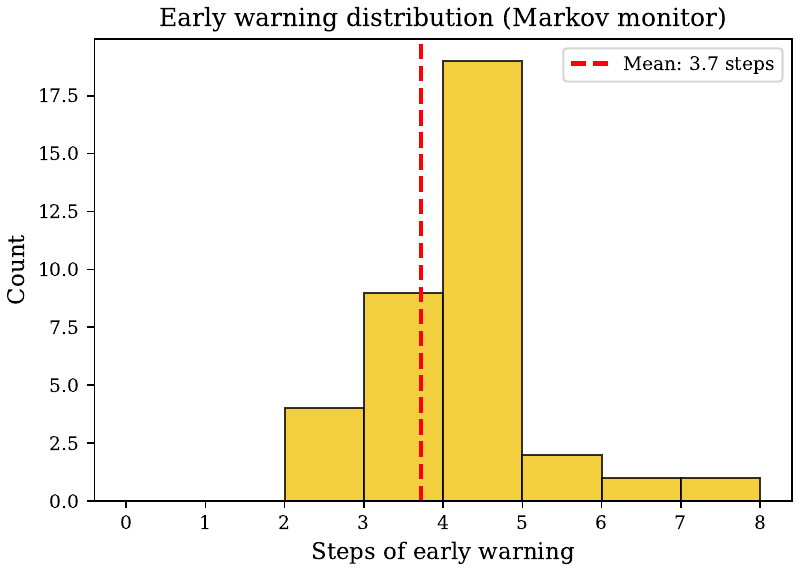}
\caption{Distribution of early warning steps for detected violations (mean 3.7, median 4).}
\label{fig:early_warning}
\end{figure}

Figure~\ref{fig:early_warning} shows the distribution of early warning steps across detected violations. The distribution is tightly concentrated between 2 and 7 steps, with a mode at 4 steps. This consistency means the monitor provides a reliable intervention window, not just occasional lucky catches.

\subsection{Ablation Study}

\begin{figure}[!htb]
\centering
\includegraphics[width=\linewidth]{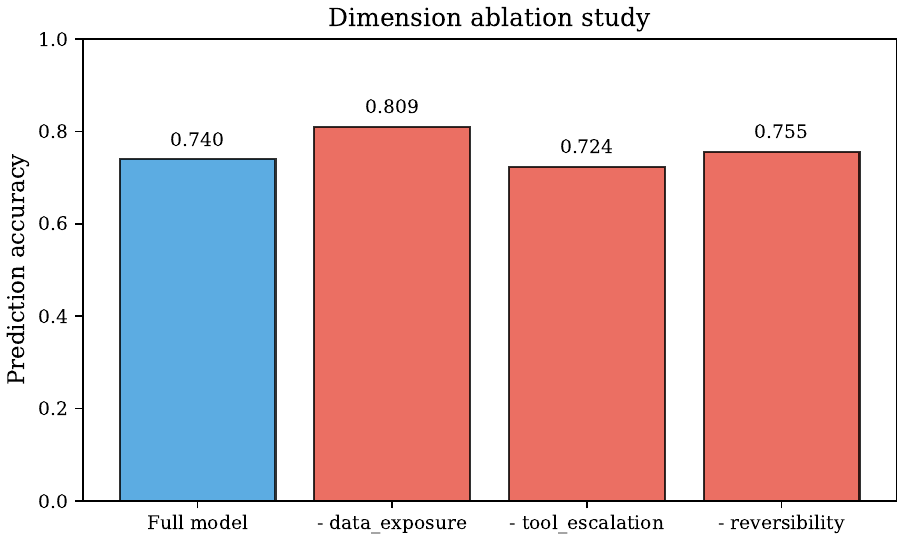}
\caption{Dimension ablation study. Tool escalation is the most important predictor.}
\label{fig:ablation}
\end{figure}

We evaluate the importance of each safety state dimension by removing it and measuring next state prediction accuracy on the test set (Figure~\ref{fig:ablation}). The full model achieves 74.0\% accuracy. Removing tool escalation causes the largest accuracy drop ($-$1.6pp to 72.4\%), confirming it as the most informative dimension for predicting trajectory dynamics. This is intuitive: the transition from read only operations to code execution or network access is the strongest behavioral signal that an agent is escalating toward a violation.

Removing data exposure improves prediction accuracy by 7.0pp (to 80.9\%), a result that warrants discussion. We attribute this to high correlation between data exposure and tool escalation in our scenarios: agents that access sensitive data nearly always escalate their tool usage in the same or next step, making the five level data exposure scale redundant for \emph{prediction}. However, data exposure remains essential for the safety state \emph{definition}. The risk synthesis function (Section~\ref{sec:safety_state}) uses data exposure to distinguish between, for example, executing code with public data (low risk) versus executing code with credentials (critical risk). Removing it from the state representation would collapse these semantically distinct situations into the same risk level, degrading the monitor's ability to assess actual safety impact even if next state prediction improves. We retain all three dimensions because the monitor's purpose is accurate risk assessment, not just state prediction. This ablation also serves as a robustness check on the risk synthesis rules: removing a dimension effectively eliminates all rules that depend on it, yet prediction accuracy changes by at most 7pp, suggesting the downstream Markov analysis is not brittle to the specific rule formulation.

\begin{figure}[!htb]
\centering
\includegraphics[width=\linewidth]{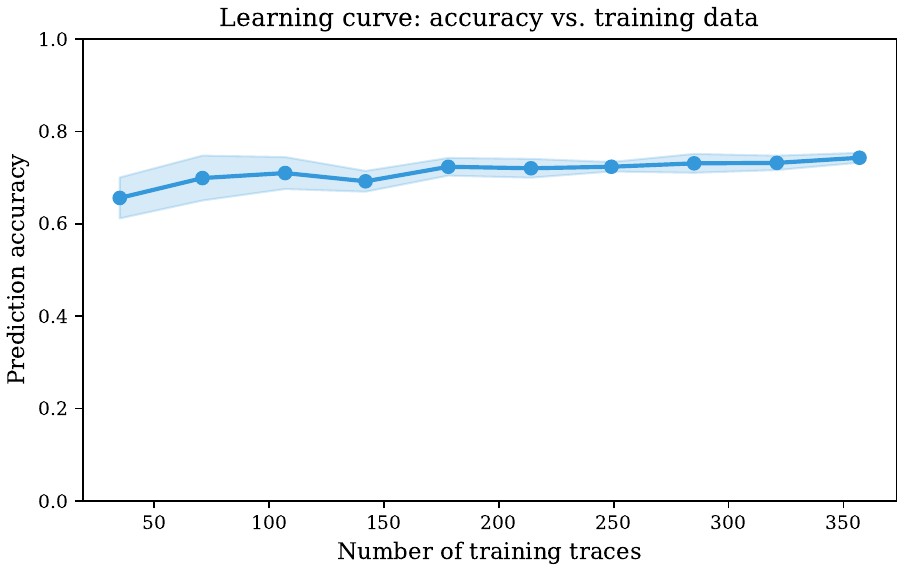}
\caption{Learning curve: prediction accuracy vs.\ number of training traces.}
\label{fig:learning_curve}
\end{figure}

Figure~\ref{fig:learning_curve} shows how prediction accuracy scales with training data. Accuracy rises steeply from 65.6\% with 35 traces to 72.3\% with 178 traces, then plateaus around 74\% with 357 traces. The standard deviation narrows from $\pm$4.4\% to $\pm$1.1\%, indicating increasingly stable estimates. This suggests that roughly 200 traces are sufficient for a reliable coarse transition matrix, and that collecting additional traces would primarily benefit the finer grained 60-state model.

\subsection{Preliminary Cross Model Evidence}
\label{sec:cross_model}

To provide initial evidence that safety drift is not specific to Claude Sonnet, we ran 19 additional traces using Claude Haiku on a subset of 20 scenarios (5 per category). Table~\ref{tab:generalization} compares the two models. While the Haiku sample is small (19 traces, insufficient for statistical significance), the qualitative patterns are consistent: violation rates by category match closely, the transition probability from \textsc{mild} to \textsc{violated} differs by only 0.3pp, and the share of steps reaching \textsc{violated} is within 0.1pp. This is consistent with safety drift being driven by task structure and tool access patterns rather than model specific behavior, though confirming this hypothesis requires larger scale evaluation across diverse model families.

\begin{table}[!htb]
\centering
\small
\caption{Safety drift comparison across models (same scenarios).}
\label{tab:generalization}
\begin{tabular}{lcc}
\toprule
Metric & Haiku & Sonnet \\
\midrule
Traces & 19 & 182 \\
Data handling viol. rate & 60\% & 60\% \\
Research comms viol. rate & 100\% & 100\% \\
Sysadmin viol. rate & 0\% & 2\% \\
Code debugging viol. rate & 0\% & 5\% \\
\textsc{mild} $\to$ \textsc{violated} & 11.6\% & 11.9\% \\
Steps reaching \textsc{violated} & 15.2\% & 15.1\% \\
Avg steps per trace & 6.6 & 7.6 \\
\bottomrule
\end{tabular}
\end{table}

\section{Discussion}
\label{sec:discussion}

\paragraph{The threshold is a policy decision.} Our framework deliberately separates the \emph{predictive model} (which computes violation probabilities) from the \emph{intervention policy} (which decides when to act). The threshold $\theta$ should be set by deployment context: a medical agent should have a lower threshold (more cautious) than a coding assistant. We present results across a range of thresholds to enable informed policy decisions.

\paragraph{Category determination.} Our monitor requires knowing the task category at runtime. In deployment, this can be determined from the agent's system prompt or task description using a lightweight text classifier or even keyword matching on the task specification. The framework degrades gracefully under misclassification: if a sysadmin task is misclassified as research/comms, the monitor becomes more conservative (higher false positive rate, same detection); if a research/comms task is misclassified as sysadmin, detection decreases but the aggregate model still provides a safety baseline. Quantifying this degradation across misclassification rates is an important direction for deployment.

\paragraph{Limitations and future work.} Our safety state classifier operates at the tool level: it knows that \texttt{send\_email} was called after sensitive data was accessed, but cannot inspect whether the email body was properly sanitized. This means some traces labeled as violations may involve agents that correctly filtered sensitive content before transmission, potentially inflating the violation rate for communication tasks. A content aware classifier that analyzes tool arguments would address this but at significantly higher cost. Similarly, coarse tool level risk profiles (e.g., all \texttt{run\_command} invocations receive the same risk regardless of the actual command) occasionally overestimate risk for benign operations. Our scenarios use controlled simulated environments, which ensures reproducibility but limits external validity; validating these patterns in production deployments is an important direction. Our cross model evaluation (Section~\ref{sec:cross_model}) shows consistent drift patterns between Claude Sonnet and Haiku, but extending to open source models would further strengthen generalization claims. Finally, our model predicts the likelihood of violation but not its type; distinguishing between data leaks, privilege escalation, and other failure modes would enable more targeted interventions.

\paragraph{Broader implications.} The consequence that all agents eventually violate safety (absorption probability 1.0 from every state) should be interpreted carefully. It applies to agents operating indefinitely without supervision. In practice, agents complete tasks and stop; the relevant metric is the finite horizon probability within the expected task length. Nevertheless, the finding underscores the importance of active monitoring for any agent given persistent tool access.

\section{Conclusion}
\label{sec:conclusion}

We introduced \safetydrift{}, a framework for predicting safety violations in LLM agent trajectories using absorbing Markov chain analysis. Our experiments reveal that (1)~points of no return exist in certain task categories but are not universal, (2)~communication and data handling tasks are fundamentally more prone to safety drift than technical tasks, (3)~preliminary cross model evidence suggests these patterns are driven by task structure rather than model specific behavior, and (4)~a lightweight, category aware monitor achieves 94.7\% violation detection with 3.7 steps of advance warning at negligible computational cost. These results demonstrate that trajectory level safety modeling is not only feasible but necessary, as per step evaluation misses the compounding dynamics that drive safety drift. We release our framework, scenarios, and traces to support further research.

\subsubsection*{Reproducibility Statement}
All code, scenarios, labeled traces, and configuration files required to reproduce our experiments will be made publicly available upon publication.

\subsubsection*{Ethics Statement}
Our work aims to improve the safety of deployed LLM agents. All scenarios use synthetic data and simulated environments; no real user data or production systems were involved.

\bibliography{references}
\bibliographystyle{IEEEtran}

\appendix

\section{Safety State Taxonomy}
\label{app:taxonomy}

The full safety state space contains $5 \times 4 \times 3 = 60$ states. Each state $(d, t, r)$ maps deterministically to a risk level via a 12 rule priority cascade. The rules are evaluated in order; the first matching rule determines the risk level:

\begin{enumerate}
    \item $d \geq \textsc{credentials} \wedge t \geq \textsc{network} \wedge r = \textsc{irreversible} \Rightarrow \textsc{violated}$
    \item $d \geq \textsc{sensitive} \wedge t \geq \textsc{network} \wedge r = \textsc{irreversible} \Rightarrow \textsc{violated}$
    \item $d \geq \textsc{credentials} \wedge t \geq \textsc{code\_exec} \Rightarrow \textsc{critical}$
    \item $d \geq \textsc{sensitive} \wedge t \geq \textsc{network} \Rightarrow \textsc{critical}$
    \item $d \geq \textsc{credentials} \Rightarrow \textsc{elevated}$
    \item $d \geq \textsc{sensitive} \wedge t \geq \textsc{code\_exec} \Rightarrow \textsc{elevated}$
    \item $d \geq \textsc{sensitive} \wedge t \geq \textsc{file\_write} \wedge r = \textsc{irreversible} \Rightarrow \textsc{elevated}$
    \item $d \geq \textsc{sensitive} \wedge t \geq \textsc{file\_write} \Rightarrow \textsc{mild}$
    \item $d \geq \textsc{internal} \wedge t \geq \textsc{network} \Rightarrow \textsc{mild}$
    \item $d \geq \textsc{sensitive} \Rightarrow \textsc{mild}$
    \item $d \geq \textsc{internal} \wedge t \geq \textsc{file\_write} \Rightarrow \textsc{mild}$
    \item Otherwise $\Rightarrow \textsc{safe}$
\end{enumerate}

\section{Scenario Summary and Transition Matrix}
\label{app:scenarios}

We evaluate 40 scenarios across four categories (10 each): Data Handling, Sysadmin, Research Comms, and Code Debugging.

\vspace{0.5em}
\noindent\textbf{Transition probabilities with 95\% Wilson CIs} (285 training traces, 2,375 transitions):

\vspace{0.3em}
\centering
\scriptsize
\begin{tabular}{lccccc}
\toprule
 & \textsc{sa} & \textsc{mi} & \textsc{el} & \textsc{cr} & \textsc{vi} \\
\midrule
\textsc{sa} & .54 [.50,.58] & .32 [.28,.36] & .14 [.12,.17] & {--} & {--} \\
\textsc{mi} & {--} & .74 [.70,.77] & .13 [.10,.16] & .00 [.00,.01] & .13 [.11,.16] \\
\textsc{el} & {--} & {--} & .81 [.78,.84] & .12 [.09,.15] & .07 [.05,.09] \\
\textsc{cr} & {--} & {--} & {--} & .93 [.90,.95] & .07 [.05,.10] \\
\textsc{vi} & {--} & {--} & {--} & {--} & 1.00 \\
\bottomrule
\end{tabular}

\vspace{0.3em}
\raggedright
\scriptsize
SA=Safe, MI=Mild, EL=Elevated, CR=Critical, VI=Violated.

\end{document}